\documentclass[11pt,twoside]{article}

\usepackage[utf8]{inputenc}
\usepackage[a4paper,twoside,bindingoffset=1cm,head=15pt,body={17cm,25cm},top=2.5cm]{geometry}
\usepackage{amsmath,amssymb}
\usepackage{graphicx}
\usepackage{color}
\usepackage{cite}
\usepackage{ifthen}
\usepackage{titlesec}
\usepackage{fancyhdr}

\titleformat{\section}{\normalfont\bfseries}{\thesection}{1em}{}
\titlespacing*{\section}{0pt}{*1.5}{1ex}

\newcounter{authorcounter}

\newcommand{\wavesauthorlist}{}
\newcommand{\wavesaddresslist}{}
\newcommand{\wavesemail}{}
\newcommand{\wavesfootnotes}{}
\newcommand{\wavesauthorpre}{}

\def\theNumberTest#1{%
  \if\relax\detokenize\expandafter{\romannumeral-0#1}\relax
    true%
  \else
    false%
  \fi
}

\newcommand{\wavesspeaker}[3][noemail]{%
    \ifthenelse{\value{authorcounter} > 1}{%
      \renewcommand{\wavesauthorpre}{, }%
    }{%
      \renewcommand{\wavesauthorpre}{}%
    }%
    \ifthenelse{\equal{#1}{noemail}}{%
      \renewcommand{\wavesfootnotes}{}
    }{%
      \renewcommand{\wavesemail}{$^\ast$Email: #1}%
      \renewcommand{\wavesfootnotes}{, \ast}
    }%
    \ifthenelse{\equal{\theNumberTest{#3}}{true}}{%
      \edef\wavesauthorlist{\wavesauthorlist%
        \wavesauthorpre{}\underline{#2}$^{#3\wavesfootnotes%
        }$%
      }%
    }{%
      \edef\wavesauthorlist{\wavesauthorlist%
        \wavesauthorpre\underline{#2}$^{\arabic{authorcounter}\wavesfootnotes%
        }$%
      }
      \edef\wavesaddresslist{\wavesaddresslist%
        \par%
        $^{\arabic{authorcounter}}$#3%
      }%
      \stepcounter{authorcounter}%
    }%
  \ignorespaces
}

\newcommand{\wavesauthor}[3][noemail]{%
    \ifthenelse{\value{authorcounter} > 1}{%
      \renewcommand{\wavesauthorpre}{, }%
    }{%
      \renewcommand{\wavesauthorpre}{}%
    }%
    \ifthenelse{\equal{#1}{noemail}}{%
      \renewcommand{\wavesfootnotes}{}
    }{%
      \renewcommand{\wavesemail}{$^\ast$Email: #1}%
      \renewcommand{\wavesfootnotes}{, \ast}
    }%
    \ifthenelse{\equal{\theNumberTest{#3}}{true}}{%
      \edef\wavesauthorlist{\wavesauthorlist%
        \wavesauthorpre{}#2$^{#3\wavesfootnotes%
        }$%
      }%
    }{%
      \edef\wavesauthorlist{\wavesauthorlist%
         \wavesauthorpre{}#2$^{\arabic{authorcounter}\wavesfootnotes%
         }$%
      }
      \edef\wavesaddresslist{\wavesaddresslist%
        \par%
        $^{\arabic{authorcounter}}$#3%
      }%
      \stepcounter{authorcounter}%
    }%
  \ignorespaces
}

\newenvironment{wavespaper}[3]{%
  \renewcommand{\wavesauthorlist}{}%
  \renewcommand{\wavesemail}{}%
  \setcounter{authorcounter}{1}%
     #2
  \twocolumn[
    \begin{center}
     \bfseries
     #1
     \bigskip

     \wavesauthorlist
     \mdseries
     \smallskip

     \wavesaddresslist
     \smallskip
 
     \wavesemail

    \end{center}%
  ]

}{%
}

\fancypagestyle{plain}{%
\fancyhead{}
}

\def\aap{{Astron.~Astrophys.}}                
\def\apj{{Astrophys.~J.\ }}                 
\def\apjl{{Astrophys.~J.~Lett.\ }}                
\def\mnras{{Mon.~Not.~R.~Astron.~Soc.\ }}             
\def\prd{{Phys.~Rev.~D\ }}        
\def\prl{{Phys.~Rev.~Lett.\ }}    
\def\pasa{{PASA\ }}               

\newcommand{\ud}{\mathrm{d}}

\begin{document}

\begin{wavespaper}{%
    Gravitational Waves from Core-Collapse Supernovae
}{%
    \wavesspeaker[b.mueller@qub.ac.uk]{Bernhard M\"uller}{Astrophysics Research Centre, School of Mathematics
    and Physics, Belfast, UK}
}{%
Vuk~Mandic}

\section*{Abstract}
Gravitational waves are a potential direct probe for the
multi-dimensional flow during the first second of core-collapse
supernova explosions. Here we outline the structure of the predicted
gravitational wave signal from neutrino-driven supernovae of
non-rotating progenitors from recent 2D and 3D simulations. We sketch
some quantitative dependencies that govern the amplitudes of this
signal and its evolution in the time-frequency domain.

\smallskip

\noindent\textbf{Keywords:}
supernovae, gravitational waves, hydrodynamic instabilities

\section{Introduction}
Core-collapse supernovae \cite{janka_12} are the explosions
of stars that begin their life with a mass of $\mathord{\gtrsim}
8M_\odot$. Such stars go through successive nuclear burning stages,
synthesizing progressively heavier elements up to the iron
group. Eventually, they build up a degenerate iron core at the
centre. As the core comes close to its effective Chandrasekhar mass,
electron capture reactions and photodisintegration of heavy nuclei
trigger its collapse.  After reaching supranuclear densities, the core
rebounds, and a shock wave is launched into the infalling outer
shells.  The shock initially stalls, and is later revived on
a time scale of hundreds of milliseconds, likely by neutrino heating,
or in rare cases of rapidly rotating progenitors by
magnetohydrodynamic effects.  Hours later, the supernova outburst
becomes visible across the electromagnetic spectrum when the shock reaches
the stellar surface.

Both observations 
and theory have shown that core-collapse
supernova explosions typically exhibit strong asymmetries
\cite{foglizzo_15}. In slowly spinning progenitors, these
asymmetries arise naturally already during the first second as neutrino
heating drives convective overturn behind the shock
\cite{herant_94,burrows_95} or from the large-scale 
``standing accretion shock instability'' (SASI)
\cite{blondin_03}.
  In fact, these
instabilities are thought to play an important supporting role in
the neutrino-driven mechanism.  Bipolar asymmetry also develops
naturally in magnetorotationally-driven explosions
\cite{burrows_07c,moesta_14}.

The electromagnetic signatures from core-collapse supernovae are
determined at later times, and thus only provide indirect clues
about the multi-dimensional flow during the operation of the
supernova ``engine'' for the first second(s) after
collapse. Gravitational waves (GWs) potentially provide a more direct way to
probe the dynamics in the supernova core.
Several works have already addressed the potential to discriminate
between different supernova mechanisms or at least facilitate
GW detection by using predicted waveforms from simulations
in conjunction with sophisticated data analysis methods
\cite{logue_12,hayama_15,gossan_16,powell_16}; and this prompts
the question whether a prospective GW detection could reveal
even more of the physics at play in supernova cores. Focusing
on GW emission by convection
and the SASI in neutrino-driven explosions, we
here review the predicted GW signatures and
sketch some of the physics that determines important signal properties.

\section{Gravitational Wave Emission in Core-Collapse Supernovae}
The link between non-spherical fluid flow in the supernova core
and GW emission is provided by the Einstein
quadrupole formula,
\begin{equation}
\label{eq:quadrupole}
h_{ij}^\mathrm{TT}
=\frac{2 G}{D c^4} \mathrm{STF}(\ddot{Q}_{ij}),
\end{equation}
which relates (to first order in the far-field limit) the
transverse-traceless components $h_{ij}^\mathrm{TT}$ of the metric
perturbations at a distance $D$ from the source
to the symmetric trace-free (STF) component of the
second time derivative of the mass quadrupole moment ${Q}_{ij}$ of a
system.
Dimensional analysis shows that the GW amplitude scales with the
kinetic energy $E_\mathrm{kin}$ of a system and an overlap factor
$\alpha_\mathrm{Q}$ that quantifies the overlap with the
quadrupole component that enters in Equation~(\ref{eq:quadrupole}),
\begin{equation}
h_{ij}^\mathrm{TT}
\sim 
\frac{2 \alpha_\mathrm{Q} G E_\mathrm{kin}}{D c^4}.
\end{equation}

Historically, the classical scenario
for GW emission in supernovae
has been that of the collapse and bounce of a rotating
iron core (see \cite{ott_08b,kotake_13} for reviews).
Here, $\alpha_\mathrm{Q}$ is non-zero from the outset
due to the rotational deformation of the core.
As the shape of the bounce signal is highly generic
\cite{dimmelmeier_07_a},
 the amplitude and frequency (which is
set by the fundamental quadrupolar oscillation
mode of the young proto-neutron star (PNS) \cite{fuller_15})
of the signal are suitable probes for inferring
the bulk parameters of the collapsing iron core:
For a Galactic supernova from a rapidly
rotating progenitor, the angular momentum of the
core may be measured within  $20\%$ accuracy
with Advanced~LIGO
\cite{abdikamalov_14}.

\section{The Stochastic Signal from SASI
and Convection} 
\label{eq:gw_conv}
Even in the absence of rotation, aspherical mass
motions in the neutrino heating  layer (or ``gain layer'') also lead to GW emission in the
later post-bounce phase due to temporal variations in the mass
quadrupole moment. Recent 2D and 3D simulations
\cite{marek_08,murphy_09,mueller_e_12,mueller_13,yakunin_16,andresen_16}
typically show several distinct phases: Shock ringing after prompt
convection leads to a low-frequency signal around $100 \, \mathrm{Hz}$
for about $50 \, \mathrm{ms}$, followed by a signal at several hundred
$\mathrm{Hz}$ with strong stochastic amplitude modulations, and
possibly a ``tail'' due to asymmetric shock expansion in the explosion
phase.
  There are still few 3D models,
which show lower amplitudes by a factor of $\mathord{\sim}
10$ \cite{mueller_e_12,andresen_16}, but the 3D models have also revealed a new
distinct low-frequency component from the
SASI at $100\text{--}200 \,
\mathrm{Hz}$ \cite{kuroda_16b,andresen_16}.

Inferring physics from the GW signal of non-rotating progenitors is
less straightforward because of its stochastic character. Without a clear relation
to physical parameters of the accretion flow onto the PNS,
it is also not obvious what range of variations should be expected
in the waveforms for different supernova progenitors.

\section{The Signal in the Time-Frequency Domain} 
However, a closer analysis of the waveforms in the time-frequency domain
and of the hydrodynamical processes behind GW emission reveal a more
ordered picture as we shall sketch here for the high-frequency component
of the signal. 2D simulations showed that this signal component originates
from the deceleration of downflows at the interface between the
heating and cooling region \cite{murphy_09}. This can be understood as
the stochastic excitation of an $\ell=2$ surface $g$-mode, which
leaves a clearly defined narrow frequency band
in wavelet spectrograms \cite{mueller_13}.
The time-dependent frequency $f_g$ of the GW emission from
this surface $g$-mode can be related to
the PNS mass $M$ and radius $R$, to
the electron antineutrino mean energy
$\langle E_{\bar{\nu}_e}\rangle$
(which is a potentially measurable proxy for the
PNS surface temperature), and the
nucleon mass $m_n$ as
\begin{equation}
  f_\mathrm{g}
  \approx
\frac{1}{2\pi}
  \frac{GM}{R^2}
\sqrt{1.1 \frac{m_n}{\langle E_{\bar{\nu}_e}\rangle}}
\left(1-\frac{GM}{Rc^2}\right),
\end{equation}
including relativistic correction terms \cite{mueller_13}.

In 3D, the picture is modified \cite{andresen_16}: The excitation of the
surface $g$-mode is less efficient because downflows in the heating
region are strongly decelerated before striking the PNS
surface and (in contrast to the 2D case) lack the rapid time
variability needed for resonant excitation of the surface $g$-mode.  The
high-frequency emission still follows $f_g$, but GW emission is now
mainly due to the overshooting of plumes from the convective region
\emph{inside} the PNS. Thus, there is a still
a well-defined structure in the time-frequency domain, but
the distance-independent quadrupole amplitudes
$A=h D$ are only $1\text{--}5 \, \mathrm{cm}$,
i.e.\ much smaller than in 2D. Mass motions in the gain
region instead excite low-frequency PNS surface
oscillations non-resonantly.

\section{Estimate of Gravitational Wave Amplitudes}
Understanding the GW amplitudes due to the excitation
of PNS surface oscillations by mass motions
in the gain region or PNS convection zone is less
straightforward. At present, we can only formulate a crude
physical model that roughly explains the 
amplitudes found in simulations and makes their
dependence on the parameters of the accretion flow
transparent without aiming for quantitative accuracy.

Using the stress formula
of \cite{blanchet_90}, 
the GW amplitude from oscillatory motions in the
PNS surface layer can be
estimated from the quadrupole component
$\rho_2$ of the density perturbations in this region,
and dimensional analysis suggests a relation
to the potential energy $E_g$ stored in the surface $g$-mode:
\begin{equation}
\label{eq:a_estimate}
  A=h D
  \sim
  \frac{G}{c^4}
  \int \delta \rho_2 \frac{GM}{r} \ud V
\sim
  \frac{G E_g}{c^4}.
\end{equation}
We expect the energy flux $L_g$ of excited waves
in the convectively stable region region
to scale with the convective luminosity $L_\mathrm{conv}$ and
Mach number $\mathrm{Ma}$ in the adjacent convective region \cite{goldreich_90}.
For stochastic excitation of the surface $g$-mode, coherence
of the forcing holds roughly over one turnover time scale $\tau$ so 
that we can relate $E_g$ to the kinetic energy $E_\mathrm{conv}$ in convection 
in the heating region or inside the PNS as follows,
\begin{equation}
\label{eq:eg}
E_g \sim \alpha\,  \mathrm{Ma}\, L_\mathrm{conv}  \tau
\sim \alpha\, \mathrm{Ma}\, (E_\mathrm{conv}/\tau)  \tau
\sim \alpha\, \mathrm{Ma}\, E_\mathrm{conv},
\end{equation}
where we have allowed for an additional factor $\alpha \lesssim 1$ quantifying
the overlap of the forcing with the spatial dependence and frequency of
the $\ell=2$ mode visible in GWs.

For excitation by convection in the gain region, we can estimate
the maximum amplitude $A_\mathrm{max}$ around the onset of the explosion
by expressing $E_\mathrm{conv}$ in terms
of the mass $M_\mathrm{gain}$ of the gain region and
the typical convective velocity. If we express the post-shock sound speed 
$c_\mathrm{s}$
in terms of the shock radius $r_\mathrm{sh}$ as
$c_\mathrm{s}=(GM/3r_\mathrm{sh})^{1/2}$ \cite{mueller_15a},
we find
\begin{equation}
  A_\mathrm{max}
  \sim
  \frac{G}{c^4}
  \alpha 
  M_\mathrm{gain} v_\mathrm{conv}^2 \mathrm{Ma}
\sim
  \frac{G}{c^4}
  \alpha 
   \frac{GM M_\mathrm{gain}}{3r_\mathrm{sh}} \mathrm{Ma}^3 .
\end{equation}
Relating the mass in the gain region
to the explosion energy $E_\mathrm{expl}$ via
the typical residual recombination energy
$\epsilon_\mathrm{rec}$ per baryon
\cite{scheck_06} of around
$5 \text{--}6 \, \mathrm{MeV}$
\cite{mueller_15b}, we further obtain
\begin{equation}
\label{eq:amax}
A_\mathrm{max}  \sim
    \frac{G}{c^4}
    \alpha
    \frac{E_\mathrm{expl}}{\epsilon_\mathrm{rec}}
    \frac{GM}{3r_\mathrm{sh}}
    \mathrm{Ma}^3,
\end{equation}
or by using
$r_\mathrm{sh} \approx 200 \, \mathrm{km}$
and a typical value of $\mathrm{Ma}^2=0.3$ at
shock revival \cite{summa_16}:
\begin{equation}
A_\mathrm{max} \sim  9 \, \mathrm{cm}  \times \alpha \times \left(\frac{E_\mathrm{expl}}{10^{51} \, \mathrm{erg}} \right).
\end{equation}
This estimate yields values in the ballpark of the GW emission from
non-resonant excitation of low-frequency surface oscillations
by convection and SASI in the 3D models of \cite{andresen_16}.
It qualitatively reproduces the trend towards stronger GW
signals from more energetic explosions seen in 2D models
\cite{mueller_13}.

The explosion energy is likely the dominant parameter in
Equation~(\ref{eq:amax}), and there is little reason to expect that
the other parameters are strongly anticorrelated to it.  With both
observations of Type~~IIP supernovae \cite{kasen_09} to a large
spread in explosion energies in ordinary (likely neutrino-powered)
supernovae by more than an order of magnitude, we can expect at least a
corresponding spread in wave amplitudes. This suggests there is still
room for somewhat stronger 3D signals from SASI and convection in the
gain region than those predicted by \cite{andresen_16}.

By contrast, the work of \cite{andresen_16} suggests that variations
in the GW signal from PNS convection stem primarily
from the overlap parameter $\alpha$ since one does not expect strong
variations in $E_\mathrm{conv}$ in this region. Estimates
of the kinetic energy $E_\mathrm{conv}$ of convective motions
in this region can be formed based on dimensional analysis
along the lines of mixing-length theory. Equating
the convective luminosity to the
diffusive core neutrino luminosity $L_\mathrm{core}$ yields
\begin{equation}
E_\mathrm{conv} \sim
\Delta M 
\left( \frac{L_\mathrm{core} \Delta R}{\Delta M}\right)^{2/3}
\sim
L_\mathrm{core}^{2/3} \Delta R^{2/3} \Delta M^{1/3}.
\end{equation}
in terms of the typical mass $\Delta M \approx 1M_\odot$
and width $\Delta R \approx 10 \, \mathrm{km}$ of
the PNS convection zone during the
pre-explosion phase. 
$L_\mathrm{core}$ does not vary considerably across
progenitors as reflected by modest variations
in the (individual) luminosity 
$L_\mathrm{\mu / \tau}$ of the heavy flavour
species \cite{oconnor_13},
which is a good proxy for the core
luminosity ($L_\mathrm{core}\approx 6 L_\mathrm{\mu /\tau}$).
With a typical value of $L_\mathrm{core}=10^{53} \, \mathrm{erg} \, \mathrm{s}^{-1}$
and $\mathrm{Ma}=0.05$, we obtain
$A\sim 1\,\mathrm{cm} \times \alpha$, again roughly in line
with \cite{andresen_16}.

\section{Outlook}

Here we have attempted to qualitatively outline some of
the relevant physical dependencies that govern the structure of the GW
signal from convection and SASI during the accretion phase in
non-rotating supernovae, underscoring recent findings
that the predicted GW signal from non-rotating neutrino-driven supernovae
contains significant physical information despite its stochastic
character. The identification of clearly defined
structures of the GW signal in the time-frequency domain has seen
further progress than can be discussed here in detail with the
identification of SASI activity via GW spectrograms
\cite{kuroda_16b,andresen_16} and of a host of modes in the
PNS in long-time simulations of potential collapsar
progenitors \cite{cerda_13}. It now remains to be seen whether these
findings can be exploited to infer physics from the prospective
detection of GWs from a supernovae, or a least facilitate such a
detection with the help of advancements in data analysis techniques.





\end{wavespaper}

\end{document}